\documentclass[useAMS,usenatbib]{mn2e}
\usepackage{float}

\usepackage{graphics}
\usepackage{graphicx}
\usepackage{verbatim}

\begin{document}

\title{Towards Laser-Guide-Stars for Multi-Aperture Interferometry: an application to the Hypertelescope}

\author[P. D. Nu\~nez, A. Labeyrie, P. Riaud]{Paul D. Nu\~nez$^{1,2}$\thanks{E-mail:paul.nunez@college-de-france.fr},
Antoine Labeyrie$^{1,2}$, Pierre Riaud $^{3}$\\
$^{1}$Coll\`ege de France, 11 place Marcelin Berthelot 75005, Paris, France\\
$^{2}$Laboratoire Lagrange. Observatoire de la C\^ote d'Azur, BP4209, Nice, France\\
$^{3}$60 rue des Bergers, 75015 Paris, France}

\maketitle

\begin{abstract}
Optical interferometry has been successful at achieving milliarcsecond resolution on bright stars. Imaging performance can improve greatly by increasing the number of baselines, which has motivated proposals to build large ($\sim 100\,\mathrm{m}$) optical interferometers with tens to hundreds of telescopes. It is also desirable to adaptively correct atmospheric turbulence to obtain direct phased images of astrophysical sources. When a natural guide star is not available, we investigate the feasibility of using a modified laser-guide-star technique that is suitable for large diluted apertures. The method consists of using sub-sets of apertures to create an array of artificial stars in the sodium layer and collecting back-scattered light with the same sub-apertures. We present some numerical and laboratory simulations that quantify the requirements and sensitivity of the technique.  
\end{abstract}

\section{Introduction}

Long-baseline optical interferometry has provided the highest angular resolution at visible wavelengths. Actual images of the stellar surfaces and circumstellar environments have been reconstructed by means of aperture synthesis (e.g. \citet{altair, zhao, millour}), which have in turn furthered our understanding of stellar astrophysics. There are proposals to improve the $(u,v)$ coverage in optical interferometry by increasing the number of apertures (e.g. \citet{labeyrie_spie}), which may lead to instruments with a light-collecting area comparable to the planned $40\,\mathrm{m}$-class telescopes. Such instruments will initially perform speckle interferometry observations, but in order to have direct imaging capabilities on rather complex sources, it will eventually be necessary to correct atmospheric wavefront distortions with adaptive optics that co-phase the interferometer. Because direct imaging is a coherent form of aperture synthesis, it is more sensitive than the incoherent combination of data from sequential baselines.

The technique of adaptive optics has undergone a rapid development in the past decades and has provided nearly diffraction-limited images with monolithic telescopes. When natural guide stars are not available, artificial sodium Laser-Guide-Stars \citep{lgs}, hereafter LGS, have enabled wavefront sensing down to a fraction of the wavelength on large telescopes (e.g. \citet{hackenberg, perrin, van_dam, rigaut}) However, the same wavefront sensing techniques cannot be applied to large ($\sim 100\,\mathrm{m}$) diluted apertures due to modest sodium layer altitude ($\sim 92\,\mathrm{km}$) and the fact that small ($\sim 10\,\mathrm{m}$) interferometric baselines can typically resolve the LGS. The idea of a LGS for optical interferometers is outlined by \citep{hlgs}, and the purpose of this paper is to further quantify the capabilities and requirements of the method. This paper focuses on the wavefront-sensing technique, on the sensitivity estimation, and on the comparison to other wavefront-sensing methods. 

The outline of the paper is as follows: Section \ref{hypertelescope} describes the Hypertelescope concept, a type of diluted optical telescope. Section \ref{hlgs} outlines a possible Hypertelescope-Laser-Guide-Star concept. Section \ref{piston} describes the wavefront sensing method, focusing only on sensing optical path differences (OPD or piston) errors between 4 sub-apertures. Section \ref{piston} also provides some sensitivity estimates and numerical simulations. Section \ref{experiment} describes an initial experiment that validates the concept in the laboratory. Section \ref{n_by_n} describes the case of a NxN array with numerical simulations. Section \ref{comparison} compares this method to other novel and related wavefront sensing methods. Sections \ref{discussion} and \ref{conclusions} provide further discussion and conclusions.

\section{The Hypertelescope} \label{hypertelescope}

Since the first two-telescope interferometer \citep{labeyrie_75}, today's larger versions have used optical delay lines. The scarce number of available delay-lines in current facilities has limited the number of optical beams that can be coherently combined. An optical interferometer with a spherical (\emph{Carlina} or Arecibo) architecture, is a new type of interferometer that will allow us to combine beams from tens to hundreds of apertures without the use of delay lines  \citep{lardiere, pedretti, riaud, labeyrie_spie}. The beam combination can be done in Fizeau mode, but as the interferometer becomes more diluted, the sampling of a large number of fringes can become difficult. A way to overcome this problem is by densifying the pupil \citep{labeyrie_1996, lardiere2, patru}. Pupil densification is a wavefront manipulation that reduces the size of the diffraction envelope and concentrates most of the light on a reduced number of fringes. A large diluted telescope with a densified pupil is also referred to as a \emph{hypertelescope}. Preliminary on-sky tests with miniature versions have been performed by \citet{pedretti} and \citet{gillet}, and the Carlina architecture has been demonstrated by \citet{carlina}. This has motivated the construction of a prototype that exploits the natural curvature of a high alpine valley \citep{labeyrie_spie}.

Possible science applications range from stellar astrophysics by directly imaging the surfaces of stars, to exoplanet science by imaging transits, to deep field galaxies and cosmology \citep{labeyrie_spie}. This interferometer will be initially used in speckle mode, which may already provide reasonably good reconstructed images of simple objects. For imaging complex and/or faint objects it is desirable to adaptively phase the light between individual sub-pupils. The hypertelescope has motivated the development of adaptive optics for diluted apertures, but other interferometers or telescopes  with segmented pupils may also possibly benefit from the technique described below.

\section{The Hypertelescope Laser-Guide-Star} \label{hlgs}

In this paper we assume that no natural guide star is available and will concentrate on a method for sensing piston differences between individual sub-apertures. This will be the dominant wave-front error when sub-apertures sizes are comparable to the Fried parameter as is the case in the ``Ubaye'' prototype \citep{labeyrie_spie}.

Conventional laser-guide star techniques cannot be used with a large ($\sim 100\,\mathrm{m}$) diluted aperture for two main reasons:
\begin{enumerate}
\item A typical artificial star of a few arc-seconds in angular size would be resolved at visible wavelengths with small ($\sim 10\,\mathrm{m}$) interferometric baselines.
\item Light rays from the star would sample a different atmospheric column as those from the artificial star, i.e. the \emph{Cone effect} \citep{tallon}. The cone effect becomes non-negligible for $\sim 10\mathrm{m}$ apertures \citep{viard}.
\end{enumerate}

\begin{figure}
  \begin{center}
    \scalebox{0.4}{\includegraphics{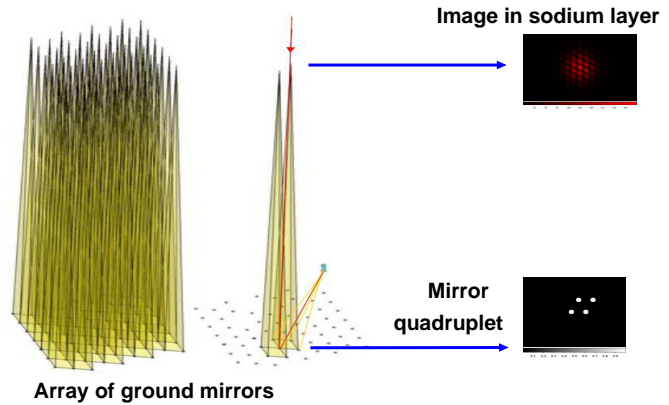}}
  \end{center}
  \caption{Quadruplets of adjacent mirrors can be used to form an array of spots in the sky. Each spot is an interference pattern, whose image in the focal plane of the hypertelescope is sensitive to wavefront errors.}\label{lgs_sketch}
\end{figure}

Our approach, described by \citet{hlgs},  is to use sub-sets of adjacent mirrors as the laser emitting optics, thus creating an array of artificial stars, where each artificial star is in fact an interference (fringe) pattern in the sodium layer (see Fig. \ref{lgs_sketch}). Each aperture sub-set can in turn be used to image the interference pattern, so that laser light passes twice through the optics. We will show in section \ref{piston} that this ``double-pass'' image can be sensitive to piston errors between sub-apertures depending on the pupil configuration as will be explained in the following sections. 

The size of each artificial star amounts to less than a meter in the sodium layer at $\sim 92\,\mathrm{km}$ altitude with $\sim 15\,\mathrm{cm}$ apertures. If this were a uniform spot, it would be resolved by the aperture and would not yield visible fringes. However, the image in the sodium layer is a set of incoherent fringes with spatial frequency corresponding to that of the sub-aperture separation. Therefore the double-pass image is a resonant convolution of intensities, and the fringes will be visible if exposure times are greater than $\sim 0.3\,\mathrm{ns}$ (corresponding to the Brownian motion time-scale, above which back-scattered light is incoherent) and less than the atmospheric coherence time of $\sim 5\,\mathrm{ms}$ at visible wavelengths. To reduce the cone effect, apertures must not be farther apart than a few meters. Even though this Hypertelescope-LGS (hereafter H-LGS) concept can solve issues (i) and (ii), it requires an array of laser guide stars, whose power requirement will turn out to be the main hard point (section \ref{laser_power}).

\section{Sensing piston errors}\label{piston}

The interference pattern formed in the sodium layer is the Point-Spread-Function (PSF) of the sub-set of apertures. The ``double-pass'' image formed at the focal plane of the diluted aperture is the convolution of the image in the sodium layer and the PSF of the reversed pupil. In Fourier space, this is the product of the auto-correlation of the quadruplet and an inverted copy of it. Therefore, the focal image of each artificial star is the Fourier transform of the squared modulus of the auto-correlation of the sub-set of apertures. The ``double-pass'' image can contain phase information as we describe below.

There is a restriction on the minimal number of apertures whose double-pass image contains phase information, e.g. the double-pass image of a pair of sub-apertures does not contain any piston information.  A triplet's ``double-pass'' image can only contain piston information if there are two redundant baselines, i.e. if the three apertures are equally spaced and collinear. However, the three adjacent sub-apertures would also have to be within a few meters of each other to reduce the cone effect. Therefore, it is reasonable to use quadruplets with two redundant baselines as shown in Fig. \ref{double-pass}. In this paper we focus on the analysis of quadruplets.

\subsection{The case of a quadruplet aperture}

\begin{figure}
  \includegraphics[scale=0.35]{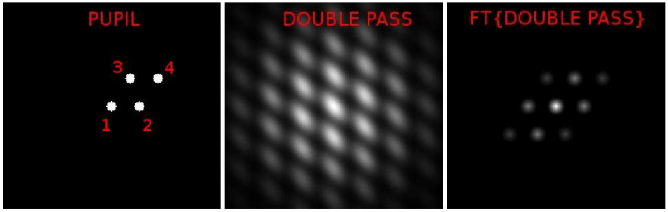}
  \caption{\label{double-pass} Each subset of apertures in the form of a quadruplet (left) can be used to form an interference pattern in the sodium layer. When the same quadruplet is used to image the interference pattern, the (middle) double-pass image is obtained. In this simulated example, there are no pistons applied. On the right we display the Fourier transform of the double-pass image.}
\end{figure}

Suppose we have a pupil with four equally illuminated sub-pupils arranged in a rhombus pattern as shown in Fig. 
\ref{double-pass}. There is an unknown atmospheric piston at each of the sub-pupils shown in Fig. \ref{double-pass}. If we assume that the wavefront is uniform across the sub-apertures, we can express the pupil function as

\begin{equation}
  P(x)=\sum_{i=1}^4\delta(x-x_i)e^{i\phi_i}. \label{pupil}
\end{equation}

To calculate the auto-correlation of $P(x)$, we define the following displacement vectors
\begin{eqnarray}
  u_1 = x_2-x_1\,&;&\,u_2 = x_3-x_1\\\nonumber
  u_3 = x_4-x_1\,&;&\,u_4 = x_2-x_3.   
\end{eqnarray}

The squared auto-correlation (optical transfer function) of the pupil is

\begin{eqnarray}
  &&\!\!\!\!\!\!\!\!\!\!\!\!\!\!|OTF|^2 = 16\delta(u)\\ \nonumber
  &+&\!\!\!\!\! 2(1+\cos(\phi_2-\phi_1-\phi_4+\phi_3))(\delta(u-u_1)+\delta(u+u_1))\\ \nonumber
  &+&\!\!\!\!\! 2(1+\cos(\phi_2-\phi_1-\phi_4+\phi_3))(\delta(u-u_2)+\delta(u+u_2))\\ \nonumber
  &+&\!\!\!\!\! \delta(u-u_3)+\delta(u+u_3)+\delta(u-u_4)+\delta(u+u_4). \label{otf}
\end{eqnarray}

\begin{figure}
  \includegraphics[scale=0.32]{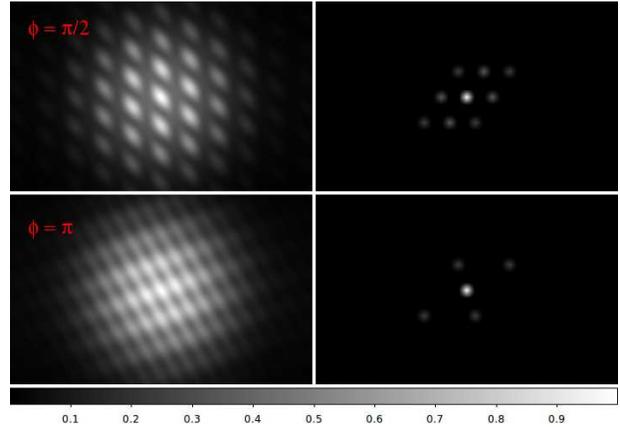}
    \caption{\label{pi2-pi}Double-pass image and it's Fourier transform for a piston error in a single sub-pupil. The top row corresponds to a delay of $\pi/2$ applied in sub-pupil 3 of the quadruplet shown in Fig. \ref{double-pass}. The bottom row corresponds to a delay of $\pi$ in the same sub-pupil.}
\end{figure}

\begin{figure*}
  \includegraphics[scale=0.6]{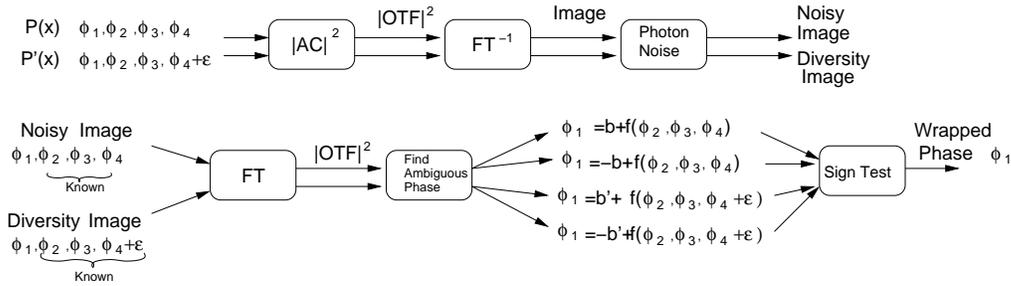}
  \caption{\label{procedure} Data simulation (above) and analysis (below) procedure for a single quadruplet aperture and a particular wavelength. The simulation receives two complex pupils: $P(x)$ and $P'(x)$, which differ by a known additive piston $\epsilon$ in one of the sub-apertures. The simulation provides two images which are fed into the analysis procedure. The analysis of each image with Eq. \ref{building_block} yields two solutions for the wrapped phase, but only one solution can be consistent with both images, so a ``sign test'' is done. The same procedure must be applied for several wavelengths in order to unwrap the phase.}
\end{figure*}

  If there are redundant baselines, such as with the rhombus pattern used in our analysis, some auto-correlation peaks will contain contributions from two redundant baselines. Therefore, in the case of the rhombus-type pupil, the squared modulus of the complex pupil's auto-correlation retains some phase information. A simulated example displaying the double-pass images with different phase delays in one of the sub-apertures is shown in Fig. \ref{pi2-pi}. The piston information can be extracted by taking the Fourier transform of the double-pass image and measuring the relative height of the central auto-correlation peak $I_0$ and the peak containing the piston information $I_1$ (second or third line of equation \ref{otf}). We see that

\begin{equation}
  \phi_2-\phi_1-\phi_4+\phi_3 =\pm \cos^{-1}{\left(\frac{8I_1}{I_0}-1\right)}. \label{building_block}
\end{equation}

Therefore, if three piston values are known from an adjacent quadruplet, the remaining piston value can be found with a sign ambiguity and modulo $2\pi$.

The phase ambiguity can be resolved by taking a separate \emph{phase diversity}\footnote{The phase diversity method is a very general method, normally used to sense wavefront errors by taking a separate image with a known additive wavefront distortion. See \citet{gonsalves} for a general description and \citet{hart} for an application to the LGS.}  image, e.g. a separate image with a known additional piston delay in one of the sub-apertures. If we denote the right side of equation \ref{building_block} by $b$, which is available from the data, a phase diversity image with a known additional piston $\epsilon$ (in aperture 2, for example), allows us to write the following two equations:

\begin{eqnarray}
  \phi_2-\phi_1-\phi_4+\phi_3 &=&\pm b \label{first}\\
  (\phi_2+\epsilon)-\phi_1-\phi_4+\phi_3 &=&\pm b'. \label{second}
\end{eqnarray}

Eq. \ref{second} above corresponds to the diversity image. Therefore, if $\phi_1+\phi_4-\phi_3$ is known, the unknown piston is $\phi_2=\pm b+\phi_1+\phi_4-\phi_3 $, and the sign ambiguity can be resolved by checking which sign is consistent with equation \ref{second}. 

So far the piston is found modulo $2\pi$ and it is desirable to find the absolute (unwrapped) piston value for polychromatic imaging. The unwrapped piston can be found by using several wavelength channels. A polychromatic LGS can be made with two lasers, one emitting at $589\,\mathrm{nm}$, and another at $569\,\mathrm{nm}$ \citep{plgs}. We investigate phase unwrapping with the use of the Sodium doublet ($589\,\mathrm{nm}$ and $589.6\,\mathrm{nm}$) and the fainter $330\,\mathrm{nm}$ line, produced from a polychromatic LGS. The doublet wavelength difference determines the maximum OPD, within which the absolute piston can be found, and is approximately , $0.6\mathrm{mm}$. However, using two very close wavelengths requires a very accurate piston measurement ($\Delta\phi\sim \lambda/200$), which motivates the need for another more separated wavelength ($330\,\mathrm{nm}$).

For each wavelength $\lambda_i$ there is a set of possible absolute OPDs

\begin{equation}
  \delta_{\lambda_i}=\frac{\phi_{\lambda_i}\lambda_i}{2\pi}+n_{\lambda_i}\lambda_i, \label{list}
\end{equation}

where the $n_{\lambda_i}$ are integers. The solution corresponds to the first overlap of the sets, and the accuracy of the solution is discussed in section \ref{sensitivity}.

\subsection{Data simulation and analysis procedure for a quadruplet}

The numerical procedure for a quadruplet consists of a data simulation stage and a data analysis stage as shown in Fig. \ref{procedure}. The data simulation receives two complex (phase sensitive) pupil configurations: $P(x)$ and a diversity pupil $P'(x)$. The two pupils differ by adding a known  piston, $\epsilon$, in one of the sub-apertures. Double-pass images are simulated by taking the inverse Fourier transform of the squared-autocorrelation of each pupil. Poisson distributed noise is finally added to each image.

The data analysis stage receives the noisy images generated by $P(x)$ and $P'(x)$, the later being the diversity image. First, we calculate the Fourier transform of each image. Next, we use equation \ref{building_block} to find two possible solutions for the wrapped phase (i.e. modulo $2\pi$) given by Eq. \ref{first} and shown as the ``Find Ambiguous Phase'' block in Fig. \ref{procedure}. A unique solution for the wrapped phase can be found by checking which solution is consistent with equation \ref{second}, which is obtained from the diversity image (``Sign Test'' in Fig. \ref{procedure}). To find the absolute OPD, we find the wrapped phase for 3 different wavelengths ($589\,\mathrm{nm}$, $589.6\,\mathrm{nm}$, $330\,\mathrm{nm}$), which provides 3 lists of the form of Eq. \ref{list}. The best overlap of these 3 lists provides the  absolute OPD, and therefore, the unwrapped phase.

\subsection{Sensitivity of a quadruplet aperture}\label{sensitivity}

\begin{figure}
  \includegraphics[scale=0.5]{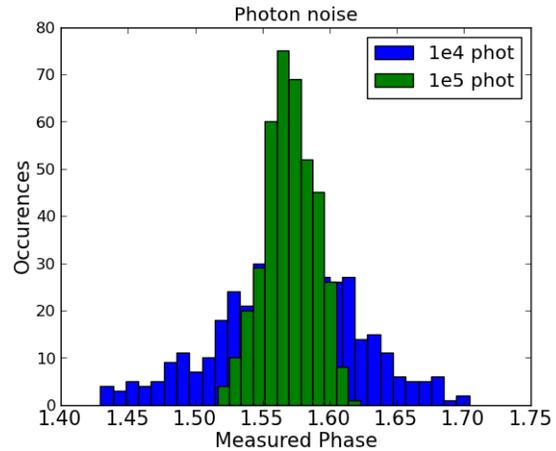}
  \caption{\label{histogram} Histogram of the occurrences of piston measurements for the cases of $10^4$ and $10^5$ photons per simulated exposure with an applied piston of $\phi=\pi/2$ and photon noise. }
\end{figure}

In this section we focus on the effect of photon noise. Monte-Carlo simulations were used to quantify the sensitivity of piston measurements in a quadruplet. Simulated double-pass images with a piston applied to one of its apertures and containing photon noise were generated. The image was then analyzed by taking its Fourier transform and using equation \ref{building_block}. The process was repeated many times to generate a simulated statistical ensemble of measurements. Fig. \ref{histogram} displays two histograms showing the number of occurrences as a function of the phase for different number of photons per double-pass image. The signal-to-noise ratio (SNR) is then the reciprocal of the standard deviation of the ensemble of measurements.

\begin{figure}
  \includegraphics[scale=0.35, angle=-90]{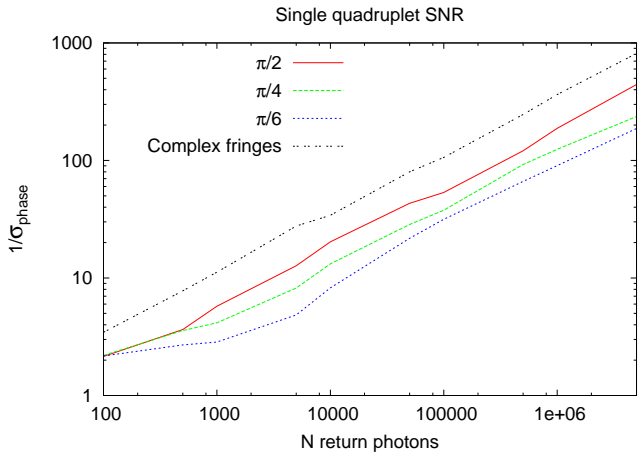}
  \vspace{0.5cm}
  \caption{\label{snr} SNR of simulated piston measurement at $589\,\mathrm{nm}$ for a single quadruplet of apertures as a function of number of photons. Each of the colored curves corresponds to a different atmospheric piston in one of the sub-apertures. The top curve is the SNR of the (hypothetical) piston measurement of the (complex) fringes in the sodium layer. }
\end{figure}

It is expected that the SNR of the piston measurement increases as a the square root of the number of photons. This is indeed the case as shown in Fig. \ref{snr}, which shows the SNR as a function of the number of photons for different atmospheric piston values. It is also worth noting that the SNR depends on the value of the atmospheric piston, where the SNR decreases as the atmospheric piston tends to $0$ or $\pi$. The fact that the SNR depends on the piston value can be seen from equation \ref{building_block} since the derivative of $cos^{-1}\beta$ tends to infinity when $\beta\rightarrow 0$ or $\beta\rightarrow \pi$. This issue should be resolvable by appropriately including additional delays when approaching a zero (or $\pi$) optical path difference, although such delay is not included in the simulations below. From Fig. \ref{snr}, it is tempting to to conclude that all the double-pass sensitivity curves tend to converge to the same SNR at low photon number, however, we note that when the OPD approaches $0$ or $\pi$, the statistical mean of each ensemble is far from the nominal (simulated) one.

The SNR of a hypothetical direct phase measurement, i.e. a direct phase measurement of fringes in the sodium layer, is shown in the curve labeled ``Complex fringes''. The SNR of a direct phase measurement is expected to be higher since the contrast of the complex fringes is higher by a factor of $\sim\sqrt{2}$ (or better) when compared to the double-pass fringes. According to Fig. \ref{snr}, the required number of detected photons per exposure for achieving a SNR of $\sim 3$ is of the order of $\sim 10^3$. However, many more photons are needed in order to unwrap the phase as we describe below.

The accuracy of phase unwrapping was investigated for the case of a quadruplet aperture via Monte Carlo simulations. Simulated double-pass images with random OPDs between $-5000\,\mathrm{nm}$ and $5000\,\mathrm{nm}$ were generated at 3 different wavelengths: $589\,\mathrm{nm}$, $589.6\,\mathrm{nm}$ and $330\,\mathrm{nm}$. The respective relative return flux for each wavelength are approximately $1$, $1/2$ and $1/60$ for the parameters used in our simulations \citep{guillet}. The probability of finding the correct unwrapped phase, that is, the correct $n_{\lambda_i}$ in Eq. \ref{list}, is shown in Table \ref{unwrap} for different amounts of detected photons. Probabilities greater than $90\%$ are found when more than $10^{6}$ photons are detected.

\begin{table}
\begin{center}
\begin{tabular}{||c||c||}\hline\hline
$N_{phot}$ & $P_{unwrapp}$\\ \hline
$10^4$ & 0.25 \\ 
$10^5$ & 0.62 \\ 
$10^6$ & 0.90\\  
$10^7$ & 0.98\\
\hline
\end{tabular}
\end{center}
\caption{\label{unwrap} For a given number of detected photons $N_{phot}$, there corresponds a probability $P_{unwrapp}$, to correctly unwrap the phase.}
\end{table}

\section{Laboratory simulations}\label{experiment}

\begin{figure}
  \includegraphics[scale=0.3]{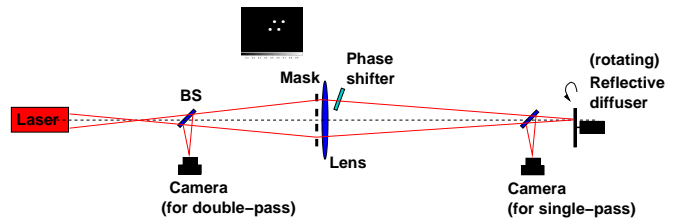}
  \caption{\label{setup} Laboratory setup of a simulation of a LGS for a diluted aperture.}
\end{figure}

For a quantitative comparison with numerical simulations, we have refined the laboratory simulations described by \citet{hlgs}, which simulate piston measurements of the wrapped phase for a single quadruplet aperture. As sketched in Fig. \ref{setup}, a $3\,\mathrm{mW},\,635\,\mathrm{nm}$ diode laser is focused to uniformly illuminate a rhombus-type mask at $72\,\mathrm{cm}$ from the focus of the laser. The mask has $2\,\mathrm{mm}$ apertures separated by $7\,\mathrm{mm}$. Light exiting the mask is focused with a lens ($f=40\,\mathrm{cm}$) on a reflective diffuser ($90\,\mathrm{cm}$ away from the lens) which is allowed to rotate (at $\sim 1\,\mathrm{Hz}$) about the optical axis. The mask and lens simulate the laser-emitting optics and the rotating reflective diffuser simulates the sodium layer. The returning light from the diffuser is imaged in the focal plane of the laser with the use of a beam-splitter (leftmost camera in Fig. \ref{setup}). The leftmost camera simulates the wavefront-sensing camera, and its exposure time needs to be longer than the speckle lifetime (of $\sim 1\,\mathrm{ms}$) to simulate an image of incoherent light emitted from the sodium layer.

To simulate the atmospheric turbulence in a controlled way, a thin film is placed in front of one of the sub-apertures, and can be tilted in order to vary the piston in a single sub-aperture. In order to know the amount of introduced piston, a separate camera (rightmost camera in Fig. \ref{setup}) images the fringes in the reflective diffuser, and the phase can be measured directly by taking the Fourier transform of these (single-pass) fringes. The phase measurement of the  ``single-pass'' fringes is much more precise (more photons) than the phase measurement of the double-pass fringes and can be used to see if the double-pass fringe measurement yields sensible results.

\begin{figure}
  \begin{center}
  \begin{eqnarray}
    \includegraphics[scale=0.25]{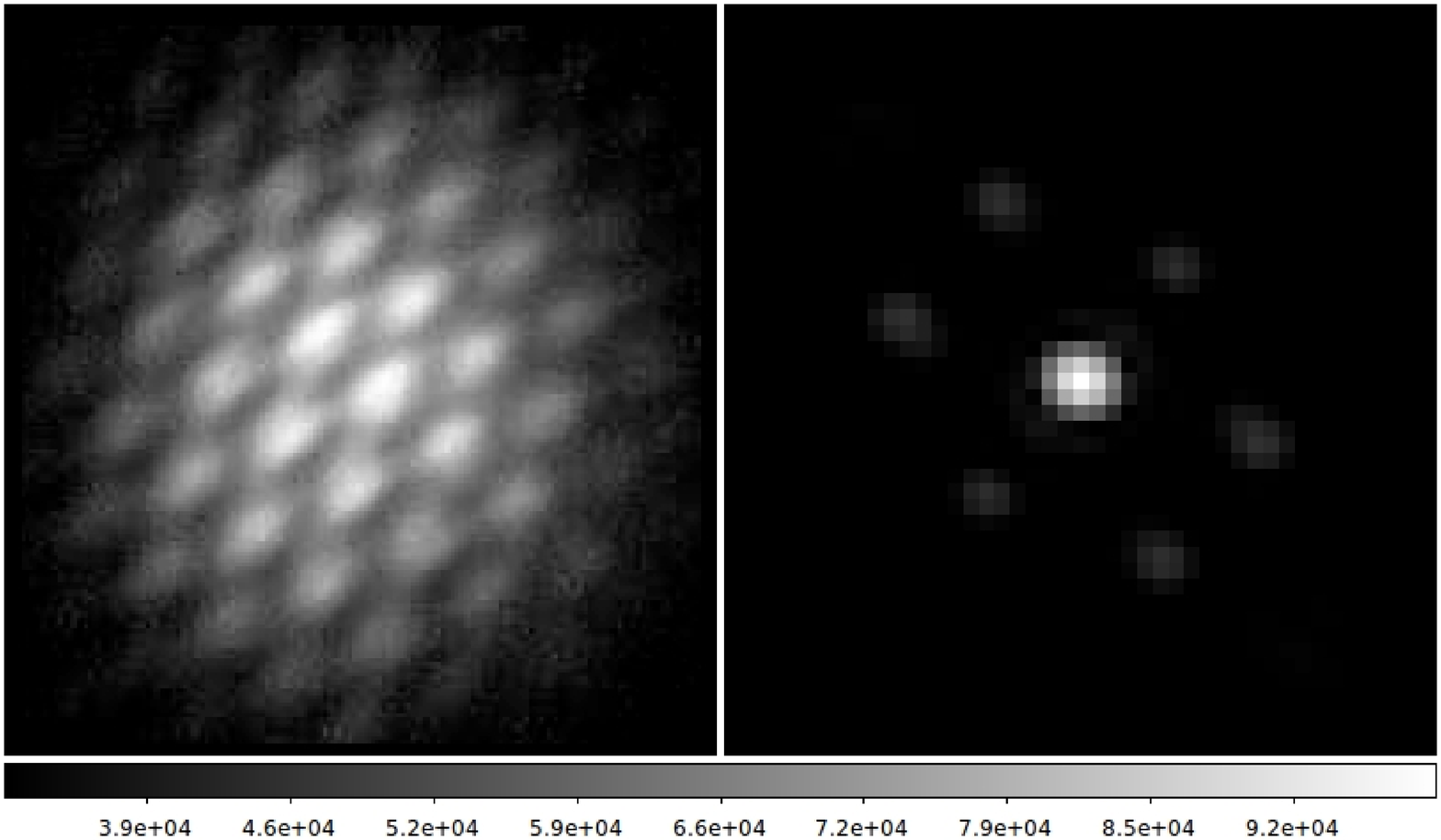}\nonumber\\
    \includegraphics[scale=0.25]{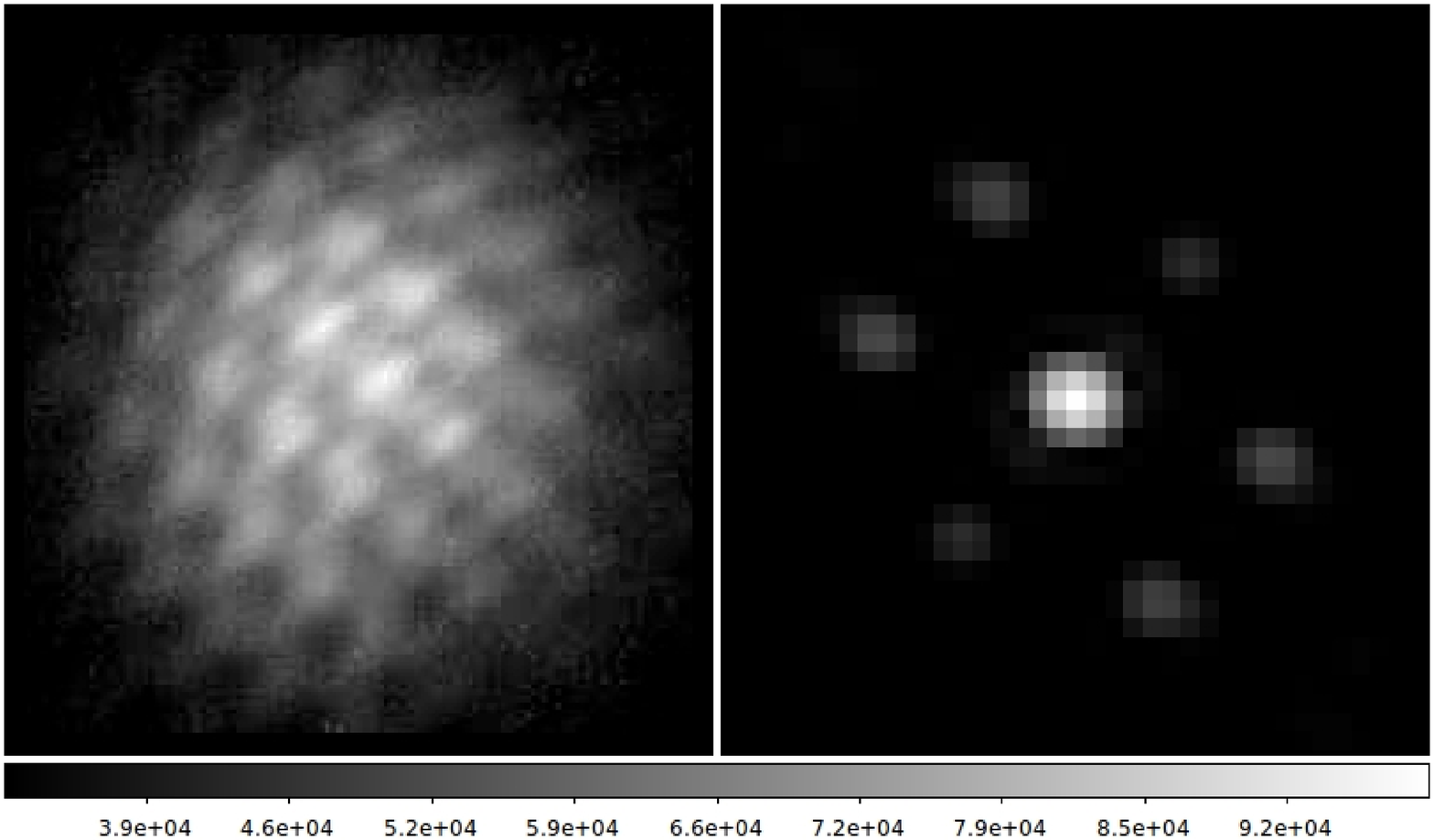}\nonumber
  \end{eqnarray}
  \end{center}
  \caption{\label{example} Double pass fringes (left) obtained in the laboratory and their Fourier transform (right). The two double pass fringes correspond to a different OPD applied to one of the apertures. The appearance of crossed fringes in the lower image as well as the change in relative intensity of the Fourier peaks evidences the phase sensitivity. }
\end{figure}

An example of two double-pass fringe images and their Fourier transform is displayed in Fig. \ref{example}. A constant background, which originates from unwanted reflections from the mask, is subtracted from the images before calculating the Fourier transforms shown in Fig. \ref{example}. The piston is estimated from each Fourier transform by using equation \ref{building_block}. Since the measurable quantity is a linear combination of piston values (eq. \ref{building_block}), there are most likely non-zero OPDs among the other three apertures, and the non-varying piston values cannot be set to zero. Therefore, we see a constant offset between the double-pass and single-pass (calibration) measurements. In Fig. \ref{comparison}, the results of an initial experiment allow us to compare the double-pass and single-pass piston measurements with a subtracted constant offset between the single-pass and double-pass piston measurements.

\begin{figure}
  \includegraphics[scale=0.35, angle=-90]{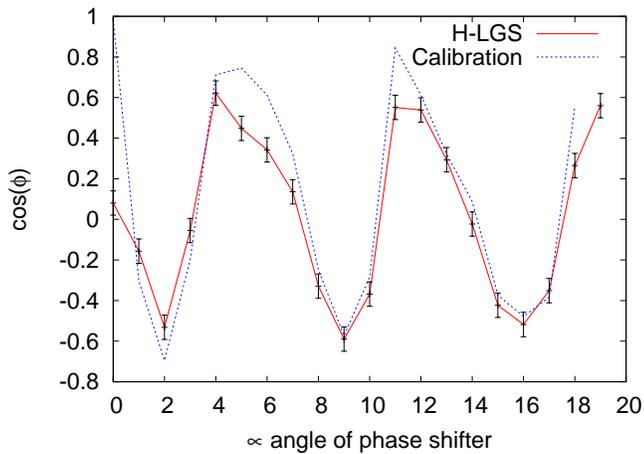}
  \vspace{0.5cm}
  \caption{\label{comparison} Piston values found from simulated LGS images (red) compared to the direct phase measurement of the fringes in the simulated sodium layer (blue). The error-bars are estimated from the statistical standard deviation of 100 measurements of the same piston.}
\end{figure}

The estimated number of detected return photons per $0.1\,\mathrm{s}$ exposure is $\sim 10^6$, which mainly accounts for the experimental geometry. According to Fig. \ref{snr} this corresponds to a theoretical SNR of $\sim 100$. The standard deviation of 100 measurements of a piston measurement of $60^{\circ}$ yields an experimental SNR closer to 10. The discrepancy is likely caused by residual background and/or having a non-uniform phase in the aperture that is in front of the thin film acting as a phase shifter. This issue deserves more attention in a future simulation experiment which is currently under development.

\section{Case of a NxN array}\label{n_by_n}

\subsection{Finding the phase across an array}

\begin{figure}
  \begin{center}
    \includegraphics[scale=0.2]{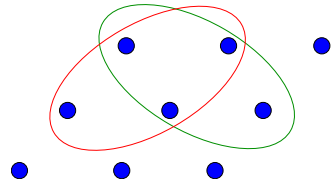}
  \end{center}
  \caption{\label{raster} A possible scheme of finding the phase across an array: Three of the phases in the leftmost quadruplet (circled in red) can be set to an arbitrary value. The analysis of it's double-pass image yields the remaining phase. The analysis of an adjacent laser-guide-star, corresponding to the adjacent quadruplet (circled in green), yields the next unknown phase.}
\end{figure}

Provided that there are several overlapping quadruplets, the phase can be found across an array as sketched in Fig. \ref{raster}: The analysis of a double-pass image of a laser-guide-star formed with a quadruplet on the edge of the array yields one of the pistons (for instance by setting the other three to zero, which ultimately results in and unknown global tip/tilt). An adjacent quadruplet with 3 overlapping apertures can be used to form an adjacent laser-guide-star, whose analysis allows us to find the next piston in the array.  The process can continue until the piston is found across the whole array, up to an unknown gross tip/tilt. All focal images should be taken simultaneously, and the phase should be found across the array  within the atmospheric coherence time ($< 5\,\mathrm{ms}$).

\subsection{Sensitivity of a NxN array}\label{n_by_n_sensitivity}

Next we estimate the required number of detected return photons per quadruplet to phase a periodic array of $N$ apertures. Simulated double-pass images are generated under realistic turbulence conditions using Van K\'arm\'an statistics\footnote{We assume a wind speed of $15\,\mathrm{m/s}$, a seeing of $0.85''$ at $550\,\mathrm{nm}$, the inner scale $l_0=1\,\mathrm{cm}$, the outer scale $L_0=20\,\mathrm{m}$ and a sampling of $3.75\,\mathrm{cm}$ per pixel. At $\lambda=589\,\mathrm{nm}$ we take the Fried parameter $r_0=15.5\,\mathrm{cm}$ and $\tau=4.5\,\mathrm{4.5\,\mathrm{ms}}$. At $\lambda=330\,\mathrm{nm}$ we take $r_0=7.2\,cm$ and $\tau=2.2\,\mathrm{\mathrm{ms}}$.}. For each array simulation we compare the simulated wavefront with the reconstructed wavefront by calculating the residual at each sub-aperture. 

\begin{figure}
  \begin{center}
  \includegraphics[scale=0.4]{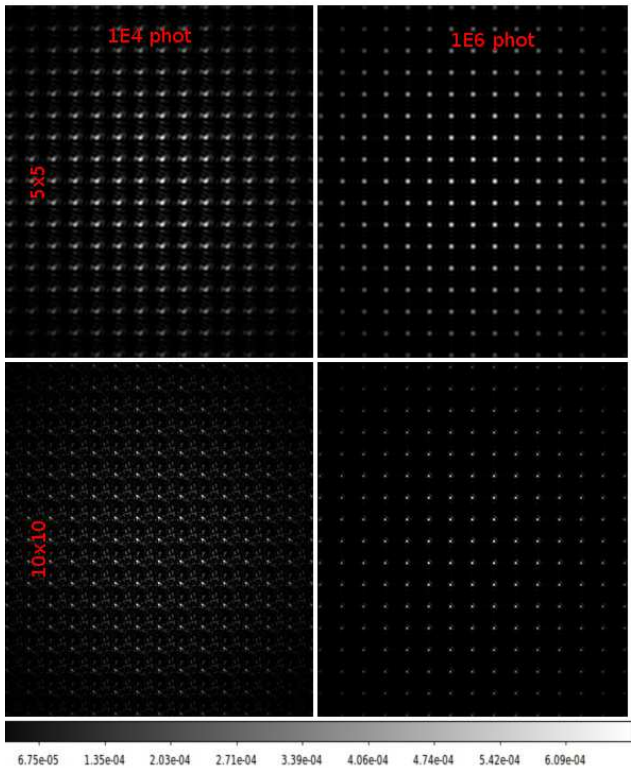}
  \end{center}
  \caption{\label{array_sim}The top row corresponds to the corrected PSF of an array of $5\times 5$ sub-pupils and different number of detected photons per aperture quadruplet, i.e. the number of detected return photons per LGS. The bottom row corresponds to an array of $10\times 10$ apertures. }
\end{figure}

\begin{figure}
  \includegraphics[scale=0.35, angle=-90]{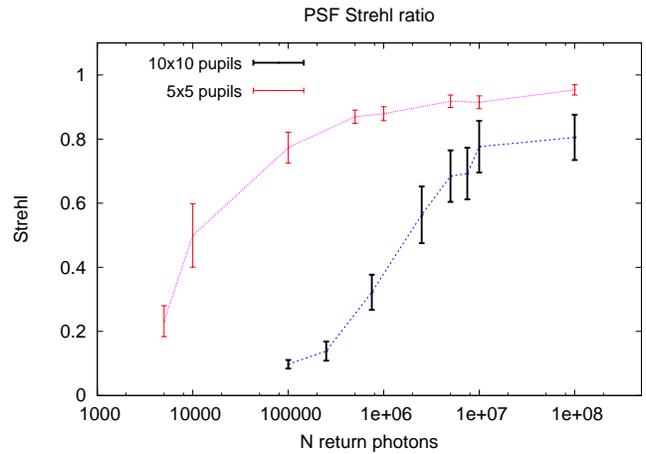}
  \vspace{0.5cm}
  \caption{\label{strehl}Strehl ratio as a function of the number of detected return photons for an array containing 25 apertures ($5\times 5$) and an array containing 100 apertures ($10\times 10$).}
\end{figure}

When all pistons are found within $2\pi$, the Strehl ratio can be estimated by finding the amount of light in the central interference peak of a monochromatic Point-Spread-Function (PSF). Fig. \ref{array_sim} displays the simulated point-spread-function (PSF) after applying piston corrections to each aperture. The PSF of an array of period $D$ and sub-aperture size $d$ is also a periodic pattern, with a period of $\lambda/(2\pi D)$ and an envelope of size $\lambda/(2\pi d)$, where $d$ is the sub-aperture size, assumed to be of the order of the Fried parameter. In the non-isopistoned case, the PSF is a speckle pattern with an envelope of size $\lambda/(2\pi d)$. Fig. \ref{strehl} shows the Strehl ratio as a function of the number of detected return photons per LGS and for two different array sizes. A minimal Strehl ratio for imaging is of the order of $\sim 0.5$, and corresponds to an average piston error of $\sim \lambda/8$ (rms). According to Fig. \ref{strehl}, this can be reached with $\sim 5\times 10^4$ detected return photons per LGS in the case of a $5\times 5$ aperture array and $\sim 5\times 10^6$ for a $10\times 10$ aperture array.

There is, however, a serious issue related to our phase-unwrapping strategy. The Strehl ratios shown in Fig. \ref{strehl} correspond to the cases when the phase was correctly unwrapped in all the sub-apertures, and the probability for this to occur is very low for the number of return photons in our simulations. With our proposed phase-unwrapping strategy, which uses 3 wavelengths, the phase-unwrapping probability at any instant of time is $\sim 50\%$ for $10^7$ detected photons per quadruplet in a 5x5 aperture array. For the case of a 10x10 array, the probability is $\sim 1\%$ for $10^7$ detected photons per quadruplet. Below $\sim 10^7$ detected photons per quadruplet, the phase-unwrapping probability is less than $1\%$. This is essentially because errors are additive across the array and the required number of detected photons per quadruplet increases in proportion to the number of quadruplets.

\subsection{Other sources of signal degradation}

In the simulations presented above, we assumed a negligible thickness of the sodium layer, and with no altitude variations as a function of time. In reality the thickness of the sodium layer is of the order of $10\,\mathrm{km}$, which turns out to be comparable to the depth of focus at $\sim 92\,\mathrm{km}$ of a a sub-set of $\sim 15\,\mathrm{cm}$ apertures. This results in a slight decrease of fringe contrast in the sodium layer, and a more detailed calculation of this effect remains to be done. If needed, the effect can possibly be corrected with range-gating techniques to separate photons received from different portions of the sodium layer and simultaneously refocus the laser light to the desired altitude.

The array of LGS will also likely produce considerable amounts of Rayleigh back-scattered light, which will produce background photon noise. Much of this can be removed in the data analysis, but the remainder will slightly affect the SNR of piston measurements. A crude estimate of this effect has been made empirically via Monte-Carlo simulations for a particular case: a constant background photon noise of $0.1$ photons per pixel to a Fizeau image whose maximal number of photons per pixel is $\sim 10$ (in the center of the field of view). In this case the SNR decreases by $\sim 10\%$.

In the simulations presented above we assume that the wavefront is coherent across an individual aperture (of $15\,\mathrm{cm}$). Degradation of the seeing may reduce the coherence across individual apertures, and in the worst case, the artificial star may resemble a speckle pattern containing fringes within each speckle. An estimate of this effect remains to be done. If needed, the usable surface area for each aperture can be reduced at the cost of a lower return flux.

The simulations presented above do not correspond to the high dilution factors\footnote{The dilution factor is defined as the ratio of the distance between sub-apertures and the sub-aperture size.} ($\sim 100$) of some planned instruments, e.g. the hypertelescope. However, numerical simulations show that, in the presence of photon noise alone, the SNR is independent of the degree of dilution as long as each fringe is sampled appropriately. High dilution factors may result in focal images which approach the sampling limit of current cameras. However, the high sampling requirement can be alleviated with pupil densification, which reduces the number of detectable fringes.

The effect of \emph{pupil densification} on the SNR was also investigated. Pupil densification can be modeled by translating the Fourier peaks of the pupil's autocorrelation to lower spatial frequencies \citep{tallon_bosc}. We followed the simulation procedure described by \citet{tallon_bosc}  to generate densified images and then introduced photon noise. A Monte-Carlo simulation for the case of $50\%$ densification was performed and no significant difference was seen compared to the non-densified case. However, we only investigated cases where the densified and non-densified images where well sampled. This is expected since the number of photons in each Fourier peak does not change.

\section{Comparison with other wave-front sensing techniques}\label{comparison}

As far as we know, there are no other proposed methods for wavefront sensing on a diluted aperture equipped with a LGS. It is still worth comparing this method with another piston sensing technique for diluted apertures with natural guide stars. When a natural guide star is available, methods such as the  ``dispersed-speckles'' technique \citep{ao_for_int} and the related ``chromatic phase diversity'' \citep{sirius} are capable of sensing pistons in an array.  The ``dispersed-speckles'' essentially takes a 3-dimensional Fourier transform of the (inverse wavelength) dispersed Fizeau image, and the positions of the Fourier peaks in this 3-D space yield the OPD for each baseline.  \citep{borkowski} have performed numerical simulations to estimate the minimum required number of photons for different applications, e.g. imaging. According to \citep{borkowski}, a dispersed-speckles data cube requires nearly 1000 photons per aperture in the case of 4 apertures to be sensitive to $\lambda/4$ OPDs. Using a similar strategy, \citep{martinache} has proposed to use a phase retrieval algorithm to measure the phase of the dispersed speckles, requiring a comparable number of photons per aperture. For the LGS scheme that we propose we also require $\sim 10^3$ detected photons to measure the wrapped phase, but our proposed phase unwrapping method requires at least $\sim 10^6$ photons as discussed in section \ref{sensitivity}. A possible advantage of our technique is that it is not necessary to look for the position of a Fourier peak to determine a piston value, which may be hidden in noise, but rather to measure the intensity of a Fourier peak at a known position determined only by the pupil configuration. 

There have been other proposals related to projecting fringe patterns in the sky with monolithic apertures, which allow measuring local phase gradients. \citet{baharav} proposed to coherently combine three beams from a single laser to create fringes in the atmosphere that are imaged by a telescope and a Shack-Hartmann wavefront sensor. \citet{love} proposed to project the pupil-plane pattern in the atmosphere and effectively create a curvature sensor \citep{roddier} by forming several images in the scattering medium. These proposals allow measuring the effect of turbulence at several layers in the atmosphere, which may also be feasible with our proposed method, but this still remains to be explored.

\section{Discussion}\label{discussion}

A LGS for interferometry requires an array of artificial stars, and the number of artificial stars is directly proportional to the number of apertures. Depending on the baseline, it may be possible to use more than 4 sub-apertures to create an artificial star, which will require a similar analysis as that described in Section \ref{piston}, and would reduce the required number of artificial stars. In any case, having several guide stars is not unheard of:  \citet{viard} proposed to use several LGSs to correct of the cone effect and the method will likely be implemented in next generation $40\,\mathrm{m}$-class telescopes. 

In Section \ref{piston} we saw that the proposed method requires redundant arrays of apertures. This requirement can be somewhat relaxed, requiring only local redundancy, by applying small ``pin-cushion'' distortions of less than the aperture size which enhance the imaging capabilities of an array. It may also be possible to perform pupil-remapping for the returning image to retain phase information of double-pass images of non-redundant configurations, although this possibility has not yet been investigated. 

It is worth noting that, as with conventional LGS systems, the global tip-tilt cannot be determined with the LGS wavefront sensor. The tip-tilt correction needs to be performed separately with a natural guide star, and the quality of this correction sets the limiting exposure time. Fortunately the isoplanatic angle is much larger for tip-tilt corrections \citep{olivier}.

\subsection{Required laser power}\label{laser_power}

The main hard point is the required laser power required for imaging purposes. A detailed simulation of sodium-layer physics to calculate the number of return photons for different types of lasers is beyond the scope of this paper. Below we provide order-of-magnitude estimates in view of the simulations made by \citep{gemini} and \citep{guillet} which include saturation effects.

\begin{itemize}
\item Requirements for a single piston measurement with a $15\,\mathrm{cm}$ quadruplet aperture: According to Fig. \ref{snr}, in order to make a $\sim 5\sigma$ measurement of a single wrapped-piston with a $5\,\mathrm{ms}$ exposure, we require $\sim 10^3$ photons. This corresponds to $10^6 \,\mathrm{phot\,s^{-1}\,m^{-2}}$ return photons, which should be achieved with a $\sim 10\,\mathrm{W}$ laser at least. However, the phase unwrapping strategy we have investigated requires at least 100 times more photons in order to obtain a repeatable measurement (see Table 1), at which the saturation of the sodium layer may be reached.

\item Requirements for co-phasing large arrays of $15\,\mathrm{cm}$ apertures with $\sim \lambda/8$ (rms) accuracy: According to Section \ref{n_by_n_sensitivity}, the required return fluxes to achieve Strehl ratios of $\sim 0.5$ with arrays of 25 and 100 apertures are $\sim 10^8 \,\mathrm{phot\,s^{-1}\,m^{-2}}$ and $\sim 10^{10} \,\mathrm{phot\,s^{-1}\,m^{-2}}$ respectively. These return fluxes suggest that it may be possible to phase an array of 25 apertures with $\sim 100\,\mathrm{W}$ lasers, and a 100 aperture array with $\sim 10\,\mathrm{kW}$ lasers. However, it is likely that saturation of the sodium layer will prevent us to obtain the high fluxes needed to phase arrays with $100$ small apertures.

\end{itemize}

In order to reduce the required return flux for $\lambda/8$ (rms) phasing, the sub-aperture size could also be increased to $\sim 1\,\mathrm{m}$, and each sub-aperture would have to be individually co-phased via other methods. In this case, the required laser power per quadruplet is of the order of $\sim 10\,\mathrm{W}$ for a 25 aperture array and $\sim 100\,\mathrm{W}$ for a 100 aperture array. If the phasing precision is further relaxed to $\lambda/4$ (rms), lower return fluxes are required, which may be sufficient for some imaging applications, although this will be further studied in a future publication. For the case of $\lambda/4$ (rms), the required laser power would also decrease by about an order of magnitude. 

We stress that the above order-of-magnitude estimates only hold when there are no phase-unwrapping errors, which is unlikely to occur with our proposed method (see Section \ref{n_by_n_sensitivity}). A different phase unwrapping strategy must be used in order do use adaptive optics on large arrays of apertures.

\section{Conclusions}\label{conclusions}
A concept for performing adaptive co-phasing of giant dilute apertures with laser guide stars has been expounded via numerical and laboratory simulations. Using subsets of apertures to form fringes in the sodium layer is a possible way to perform wavefront sensing with diluted apertures. This solves the problem of resolving the artificial star since the same subset of apertures is used to form the artificial star/fringes and to perform wavefront sensing by re-imaging the fringes, which contain OPD information. The use of several LGSs solves the cone effect problem, as done in conventional LGS methods, although the residual cone effect must be further studied. Preliminary experimental efforts validate the technique in the case of a single LGS, and with SNRs comparable to those predicted by simulations. Aside from all the engineering feats that must be undertaken, we have encountered a couple of related problems: our proposed phase-unwrapping method requires too many photons, and we are currently working on a solution for a future paper. Another problem for implementing an LGS on diluted apertures is the need of many very powerful lasers. 

\section{Acknowledgments}
We would like to thank J.B. Daban for his help with laboratory logistics. We also would like to thank F. Millour and W. Dali-Ali for useful comments and discussions.

\end{document}